\documentclass[pra,twocolumn]{revtex4}

\usepackage{bm, graphicx, amsmath, amsfonts,curves}
\usepackage[mathscr]{eucal}

\newcommand{\ket}[1]{{\left|{#1}\right\rangle}}
\newcommand{\bra}[1]{{\left\langle{#1}\right|}}
\newcommand{\braket}[3][]{{\left\langle{#2}{#1|}{#3}\right\rangle}}
\DeclareMathAlphabet{\vecfont}{OT1}{cmr}{bx}{it}
\renewcommand{\vec}[1]{\vecfont{#1}}
\newcommand*{\vsigma}{\boldsymbol{\sigma}}
\newcommand*{\vtheta}{\boldsymbol{\theta}}
\newcommand*{\Chi}{\raisebox{0.3ex}{\large$\chi$}}
\newcommand*{\ts}{\rule{0.1em}{0ex}}  
\newcommand*{\vts}{\rule{0.05em}{0ex}}  
\newcommand*{\SIC}{SIC\rule{0.2em}{0ex}POVM}
\newcommand{\Eq}[1]{Eq.~(\ref{#1})}
\newcommand*{\column}[2][@{\,}c@{\,}]%
      {{\left(\begin{array}{#1}#2\end{array}\right)}}
\newcommand*{\repr}{\mathrel{\widehat{=}}}  

\DeclareMathOperator{\trace}{Tr}
\newcommand*{\tr}[2][]{\trace_{#1}\!{\left(#2\right)}} 
\DeclareMathAlphabet{\listfont}{T1}{cmr}{bx}{sl}  
\newcommand*{\lst}[1]{\listfont{#1}}

\newcommand*{\expect}[2][]%
   {\mathbb{E}{\left(\vphantom{#1|}#2\right)}}
\newcommand*{\magn}[2][\big]%
  {\mathopen{\boldsymbol{#1|}}{#2}\mathclose{\boldsymbol{#1|}}}

\begin{document}

\title{State learning from pairs of states}
\author{Pranjal Agarwal,$^{1,2}$ Nada Ali,$^{1,2}$ Camilla Polvara,$^{2,3}$
  \mbox{Martin-Isbj\"{o}rn Trappe},$^{4}$ Berthold-Georg Englert,$^{5,4,6}$
  and Mark Hillery$^{1,2}$}
\affiliation{\rule{0pt}{16pt}%
  $^{1}$Department of Physics and Astronomy, %
  Hunter College of the City University of New York, %
  695 Park Avenue, New York, NY 10065 \\ 
  $^{2}$Physics Program, Graduate Center of the City University of New York, %
  365 Fifth Avenue, New York, NY 10016 \\
  $^{3}$College of Staten Island of the City University of New York, %
  2800 Victory Blvd., Staten Island, NY 10314 \\ 
  $^{4}$Centre for Quantum Technologies, National University of Singapore, %
  Singapore 117543, Singapore \\
  $^{5}$Key Laboratory of Advanced Optoelectronic Quantum Architecture %
  and Measurement of the Ministry of Education, School of Physics, %
  Beijing Institute of Technology, Beijing 100081, China \\
  $^{6}$Department of Physics, National University of Singapore, %
  Singapore 117542, Singapore} 

\begin{abstract}
Suppose you receive a sequence of qubits where each qubit is guaranteed to be
in one of two pure states, but you do not know what those states are.
Your task is to determine the states.
This can be viewed as a kind of quantum state learning---or quantum state
estimation.
A problem is that, without more information, all that can be determined is the
density matrix of the sequence and, in general, density matrices can be
decomposed into pure states in many different ways.
To solve the problem, additional information, either classical or quantum, is
required.
We show that if an additional copy of each qubit is supplied---that is, one
receives pairs of qubits, both in the same state, rather than single
qubits---the task can be accomplished. 
This is possible because the mixed two-qubit state has a unique decomposition
into pure product states.

For illustration, we simulate numerically the symmetric, informationally
complete measurement of a sequence of qubit pairs and show that the unknown
states and their respective probabilities of occurrence can be inferred from
the data with high accuracy.
Finally, we propose an experiment that employs a product measurement and
can be realized with existing technology, and we demonstrate how the data tell
us the states and their probabilities.
We find that it is enough to detect a few thousand qubit pairs.
\end{abstract}

\date[Posted on the arXiv on ]{May 6, 2025}

\maketitle

\section{Introduction}
Suppose we have a collection of data, about which we have very little
information, and we are interested in learning something about it.
If the data is classical, an example of such a problem is unsupervised machine
learning.
In this scenario, the objective is to classify data into clusters with the idea
that the data within a cluster are related.
There is no training phase in which sample data with their classifications are
provided; in the unsupervised case, there are only the data to work with.
Quantum algorithms have been applied to obtain speedups of the unsupervised
learning of classical data \cite{Brassard,Lloyd2,Svore,kerenidas}
(for reviews of quantum machine learning, see \cite{schuld}).
In these works, the classical data are converted into quantum states, which can
then be processed by a quantum computer. 

Quantum learning---or quantum estimation---is related to classical machine
learning, but because the objects to be learned are quantum, new elements come
into play.
One can learn a number of different quantum objects, unitary operators
\cite{bisio}, and measurements \cite{sedlak,guta,sentis,fanizza}, for example.
In many cases, there is a training set.
In the case of a unitary operator, one is allowed a certain number of uses of
the operator, and in the case of a measurement, one is given examples of the
states one wants the measurement to distinguish.
What we want to do here is to see what can be done in the case of learning a
set of unknown quantum states, in particular determining what states are in
the set.
In most approaches to quantum unsupervised machine learning, one has access to
unitary operators that produce the data by acting on a reference quantum state
\cite{kerenidas}.
What, however, can be done if this is not the case and one has access only to
the raw data, that is, just the quantum systems themselves?
The first treatment of this kind of quantum  learning was given in
\cite{bagan}.
There one is given a sequence of $N$ particles, each in one of two unknown
states, $\ket{\psi_{0}}$ and $\ket{\psi_{1}}$, and one wants to determine
the sequence.
For any individual qubit in the sequence, you do not know which state it is
in.  The output of this procedure is classical, a sequence of $0$s and $1$s,
corresponding to the labels of the states, of length $N$ that is the best
guess for the sequence of states. 

A second approach was taken in \cite{hillery}.
There the setup was the same as in \cite{bagan}, but the objective was to use
the data to construct a positive-operator-valued measure (POVM) that would
distinguish the two states.
The fundamental problem is that all one can measure is the density matrix 
of the ensemble that describes the sequence,
\begin{equation}\label{eq:0}
  \rho^{\ }_{\textrm{1qb}}
  =p_{0}\ket{\psi_{0}}\bra{\psi_{0}}+p_{1}\ket{\psi_{1}}\bra{\psi_{1}}\,,
\end{equation}
if $\ket{\psi_{0}}$ appears with probability $p_{0}$ and $\ket{\psi_{1}}$
appears with probability $p_{1}$.
A rank-two mixed state density matrix can be decomposed into a sum of pure
states in many different ways.
That means that additional information, either classical or quantum, is
required in order to determine $\ket{\psi_0}$ and $\ket{\psi_{1}}$.
Several examples of additional classical information were explored in
\cite{hillery}.
In one example, it was specified that the two states lie on a known circle of
the Bloch sphere, e.g., the intersection of the $x,z$ plane and the sphere,
and that ${p_{0}=p_{1}=1/2}$.
Under those highly restrictive conditions,
it is possible to construct a POVM that will discriminate the states.

A simple example of how extra classical information allows one to determine
which of two ensembles with the same density matrix one has is the following
\cite{Peres:93,paris}.
We have $2N$ spin-1/2 particles, which are promised to be in one of two
ensembles.
In the first, $N$ spins point in the $+x$ direction and $N$ point in the $-x$
direction, and in the second, $N$ spins point in the $+z$ direction and $N$
point in the $-z$ direction.
These ensembles are described by the same single-particle density matrix, but
they can be distinguished with high probability by measuring all the particles
in the $z$ direction.
If you find that $N$ of the particles point in the $+z$ direction, the
ensemble is with high probability the second, and if you find that some number
other than $N$ particles point in the $+z$ direction, the ensemble is
definitely the first.

In this paper, we are going to use a model similar to that employed in
\cite{bagan,hillery}, but our objective will be to determine the states.
Our main tool will be a form of state tomography.
This is a form of state learning \cite{anshu}.
In state learning, one receives many copies of the state to be learned and
this state is guaranteed to be a member of a certain set of states.
One then performs measurements on the copies, and the result is a sufficiently
accurate description of the state.
The difference in our case is that there are two states, not one, that one is
trying to learn, and the copies are scrambled; any given copy could represent
either of the quantum states, and you don't know which.
In addition, the two states could be any pure states, and you don't know with
what probability they occur.
This task sounds formidable but, as we shall see, if the states are received
in pairs, where the members of each pair are identical, tomography can be
applied to determine the states.

\section{Pairs}
In this section, we will show that if the states are received in pairs, the
situation is much improved, and all you have to know is that the two-qubit
density matrix is an incoherent superposition of two pure product states.
Receiving an extra copy of each state means that extra quantum information is
being provided.
To be more specific, consider the following scenario.
You are sent a stream of pairs of qubits, and each pair is in the state
$\ket{\psi_{0}}\otimes\ket{\psi_{0}}$ or
$\ket{\psi_{1}}\otimes\ket{\psi_{1}}$ with 
\begin{align}
\ket{\psi_{0}} & =  a_{0}\ket{0}+ a_{1} \ket{1}\,, \nonumber \\
\ket{\psi_{1}} & =  b_{0}\ket{0}+ b_{1} \ket{1}\,,
\end{align}
where $\ket{0}$, $\ket{1}$ are single-qubit orthonormal kets that serve as
the reference basis (``computational basis'').
The $\ket{\psi_{0}}$ pair occurs with probability $p_{0}$ and the
$\ket{\psi_{1}}$ pair occurs with probability ${p_{1}=1-p_{0}}$.
You do not know what $\ket{\psi_{0}}$ and $\ket{\psi_{1}}$ are, and for any
given pair, you do not know which kind of pair it is.
You also do not know $p_{0}$ and $p_{1}$.
The task is to find $\ket{\psi_{0}}$, $\ket{\psi_{1}}$, $p_{0}$, and
$p_{1}$.

One can then, for example, use this to construct a POVM to discriminate
between the different types of pairs from a subset of the pairs and then use
it to discriminate the remaining pairs.
As only two states are involved, the purpose-appropriate qubit POVM will be
used for the pair states $\ket{\psi_0}\otimes\ket{\psi_0}$ and
$\ket{\psi_1}\otimes\ket{\psi_1}$, since their distinguishability is larger
than that of the single-qubit states.
For instance, there are the POVMs for unambiguous discrimination
\cite{Chefles:98} or for extracting the accessible information
\cite{Keil:24}. 

The ensemble we are looking at is described by the density matrix
\begin{equation}
  \rho=p_{0}\ket{\psi_{0}}\bra{\psi_{0}}\otimes\ket{\psi_{0}}\bra{\psi_{0}}
  +p_{1}\ket{\psi_{1}}\bra{\psi_{1}}\otimes\ket{\psi_{1}}\bra{\psi_{1}}\,.
\end{equation}
The density matrix and the knowledge that it is composed of two pure
two-qubit product states is the only information to which we have access.
We will now show that this is sufficient to find the states and probabilities.

To begin with, we note that $\ket{\psi_{0}}\otimes\ket{\psi_{0}}$
and $\ket{\psi_{1}}\otimes\ket{\psi_{1}}$ span a two-dimensional subspace in the
four-dimensional ket space of the qubit pair; the subspace is analogous to the
Bloch sphere of a qubit.
In the corresponding Bloch ball,
the separable mixed states are located on a line that connects the points for
$\ket{\psi_{0}}\otimes\ket{\psi_{0}}$ and
$\ket{\psi_{1}}\otimes\ket{\psi_{1}}$ on the sphere;
see \cite{Sanpera+2:98,E+M1,E+M2} and, in particular, Fig.~2.2 in \cite{E+M2}.
Once we learn $\rho$ from the data, its range is the Bloch sphere and hence we
know the line of separable states and their endpoints.

More specifically, $\rho$ has support in the three-dimen\-sional symmetric
subspace of two qubits, the triplet sector, and is of rank two.
That means that there is a direction $\ket{\xi}$ in the triplet sector
that is orthogonal to $\rho$, i.e., ${\rho\ket{\xi}= 0}$.
Set 
\begin{equation}\label{xi}
  \ket{\xi} = c_{00}\ket{00} + c_{01}\bigl(\ket{01}+\ket{10}\bigr)
  + c_{11} \ket{11}\,,
\end{equation}
and let's see what being orthogonal to $\ket{\psi_{0}}\otimes\ket{\psi_{0}}$
implies.
The orthogonality condition is
\begin{equation}
  (a_{0}^{\ast})^{2} c_{00} + 2a_{0}^{\ast}a_{1}^{\ast} c_{01}
  + (a_{1}^{\ast})^{2} c_{11} = 0.
\end{equation}
Dividing both by $(a_{1}^{\ast})^{2}$ and setting
${z=a_{0}^{\ast}/a_{1}^{\ast}}$, we find
\begin{equation}
\label{quad}
c_{11} + 2c_{01}z + c_{00} z^{2} =0 .
\end{equation}
An identical equation holds for $\ket{\psi_{1}}\otimes\ket{\psi_{1}}$ except
that in that case $z$ is $b_{0}^{\ast}/b_{1}^{\ast}$.
The above equation has two solutions for $z$, one corresponding to
$\ket{\psi_{0}}$ and the other to $\ket{\psi_{1}}$.
That means that if we know $\ket{\xi}$, then we know both $\ket{\psi_{0}}$ and
$\ket{\psi_{1}}$.
That suggests that one way to proceed is to find a way to determine
$\ket{\xi}$.  This can be done by finding $\rho$ by performing state
tomography and then finding the solution to ${\rho\ket{\xi} = 0}$.
One can also find explicit formulas for
$\ket{\psi_{0}}$ and $\ket{\psi_{1}}$ in terms of $\rho$, which we shall
proceed to do. 

Note that the argument we just used can be easily generalized.
For example, suppose that our sequence consists of three rather than two qubit
states, and we receive not pairs but trios,
where all the states within a given trio are guaranteed to be the same.
The overall state of each trio lies in the symmetric subspace of the space of
three qubits---the space with total spin $3/2$, if we regard each qubit as a
spin-1/2 object---which is
four-dimensional, while the density matrix for the trios is of rank three.
That means there is a ket in the symmetric subspace that is annihilated by all
three of the states and, consequently, by the trio density matrix.
In analogy to the two-state case, this ket will lead to a cubic equation whose
solutions will yield the three states.
Clearly, the argument can be carried further to $N$ states, which would
require receiving identical $N$-tuplets. 

It is also possible to generalize this procedure to two states in a
$d$-dimensional space \cite{anon}.
Initially you conduct single-qubit tomography
to find the single-particle density matrix, which will be of rank two.
You then find its two eigenstates, call them $\ket{0}_{d}$ and $\ket{1}_{d}$.
The states we are trying to find lie in the two-dimensional subspace spanned
by  $\ket{0}_{d}$ and $\ket{1}_{d}$, and the problem has been reduced to one
of effective qubits.
One can then apply the reasoning in the previous paragraphs but substituting
$\ket{0}_{d}$ for $\ket{0}$ and $\ket{1}_{d}$ for $\ket{1}$. 

Let us now return to the case of two qubit states and find explicit formulas
for the states and probabilities.
If the Bloch vectors for $\ket{\psi_{0}}$ and $\ket{\psi_{1}}$ are $\vec{a}$
and $\vec{b}$, respectively, then the two-qubit pair density matrix for
$\ket{\psi_{0}}$ is 
\begin{align}\label{rho0}
  \rho_{0}
  & =\frac{1}{2}\bigl(I_2+\vec{a}\cdot\vsigma\bigr)
     \otimes\frac{1}{2}\bigl(I_2+\vec{a}\cdot\vsigma\bigr)\nonumber\\
  &= \frac{1}{4}\Bigl( I_{4}
       + \vec{a}\cdot \bigl(\vsigma^{(1)}+\vsigma^{(2)}\bigr)
       +  \vsigma^{(1)}\cdot\vec{a}\vts\vec{a}\cdot\vsigma^{(2)}\Bigr)\,,
\end{align}
and similarly for $\ket{\psi_{1}}$ but with $\vec{a}$ replaced by $\vec{b}$.
Here, $I_{d}$ is the $d\times d$ identity matrix,
$\vsigma$ is the generic vector of Pauli matrices,
$\vsigma^{(1)}=\vsigma\otimes I_2$ is that for qubit~1, and
$\vsigma^{(2)}=I_2\otimes\vsigma$ is that for qubit~2.
 The ensemble  pair density matrix is
\begin{align}\label{genrho}
  \rho
  & = p_{0}\rho_{0} + p_{1}\rho_{1}\nonumber\\
  &=\frac{1}{4}\Bigl( I_{4}
       + \vec{s}\cdot \bigl(\vsigma^{(1)}+\vsigma^{(2)}\bigr)
       +  \vsigma^{(1)}\cdot\vec{C}\cdot\vsigma^{(2)}\Bigr)\,, 
\end{align}
where
\begin{align}\label{gensC}
  \vec{s}
  &=\bigl\langle\vsigma^{(1)}\bigr\rangle
  =\bigl\langle\vsigma^{(2)}\bigr\rangle=p_0\vec{a}+p_1\vec{b}\,,
  \nonumber\\
  \vec{C}
  &=\bigl\langle\vsigma^{(1)}\vts\vsigma^{(2)}\bigr\rangle
    =p_0\vec{a}\vts\vec{a}+p_1\vec{b}\vts\vec{b}\,.
\end{align}
Note that $\vec{a}\vts\vec{a}$ is the dyad with matrix elements $a_ja_k$, and
similarly for $\vec{b}\vts\vec{b}$, and the dyad $\vec{C}$ has the
matrix elements
${C_{jk}=\langle\sigma^{(1)}_j\sigma^{(2)}_k\rangle%
  =\langle\sigma_j\otimes\sigma_k\rangle%
  =p_{0}a_{j}a_{k}+p_{1}b_{j}b_{k}}$;
as is characteristic for mixed triplet states,
$\vec{C}$ is symmetric, ${C_{jk}=C_{kj}}$, and has unit trace,
${\sum_jC_{jj}=1}$.

Full tomography of the qubit pairs will provide us with the vector
$\vec{s}$ and the dyad $\vec{C}$, and we demonstrate now how knowledge
of these quantities can be converted into knowledge of $\vec{a}$,
$\vec{b}$, $p_{0}$, and $p_{1}$.
In the first step, we find
\begin{equation}
  \vec{C}-\vec{s}\vts\vec{s}=p_0p_1(\vec{a}-\vec{b})(\vec{a}-\vec{b})\,;
\end{equation}
if ${\vec{C}-\vec{s}\vts\vec{s}=0}$, the source emits only one state and we
are done.
Otherwise, ${\vec{s}^2<1}$
and $(\vec{C}-\vec{s}\vts\vec{s})\big/(1-\vec{s}^2)$
projects on the direction of ${\vec{a}-\vec{b}\neq0}$.
We remove the component parallel to ${\vec{a}-\vec{b}}$ from
$\vec{s}$ and obtain
\begin{equation}
    \vec{s}'
    =\frac{\vec{s}-\vec{C}\cdot\vec{s}}{1-\vec{s}^2}
    =\frac{1}{2}(\vec{a}+\vec{b})\,.
\end{equation}
Then, $(p_0-p_1)^2$ and $\vec{a}-\vec{b}$ are available from
\begin{equation}
  \frac{(\vec{s}-\vec{s}')^2}{1-{\vec{s}'}^2}=(p_0-p_1)^2\,,
  \quad\frac{\vec{s}-\vec{s}'}{p_0-p_1}=\frac12(\vec{a}-\vec{b})\,,
\end{equation}
and we arrive at
\begin{equation}\label{ab}
    \vec{a}=\frac{\vec{s}-2p_1\vec{s}'}{p_0-p_1}\,,\quad
  \vec{b}=\frac{2p_0\vec{s}'-\vec{s}}{p_0-p_1}\,,
\end{equation}
provided that ${p_0\neq p_1}$.
If ${p_0=p_1=\frac{1}{2}}$, when ${\vec{s}'=\vec{s}}$,
we identify ${\vec{a}-\vec{b}}$ as the eigenvector of
${\vec{C}-\vec{s}\vts\vec{s}}$ with the eigenvalue
$\frac{1}{2}(1-\vec{a}\cdot\vec{b})$.

In the following sections we shall discuss two schemes for the state
tomography that provides data from which one can estimate $\rho$ and thus
$\vec{s}$ and $\vec{C}$.
Section \ref{sec:SIC} deals with the \SIC\ (Symmetric Informationally Complete
POVM) in the triplet sector;
see \cite{ziman}, for example, for properties of \SIC{}s.
The high symmetry of the \SIC\ is attractive and facilitates the analysis but
we do not know how to implement the \SIC\ in the laboratory.
By contrast, the tetrahedron POVM of Sec.~\ref{sec:tetra} can
definitely be realized with existing technology.

\section{\SIC}\label{sec:SIC}
\subsection{The measurement}
In the triplet sector, the symmetric subspace of two qubits, we
expand the kets in the basis used in \Eq{xi},
\begin{equation}
  \ket{v}=a_0\ket{00}+a_1\ket{11}
  +a_2\frac{\ket{01}+\ket{10}}{\sqrt{2}}
  \repr\column{a_0 \\ a_1 \\ a_2}\,.
\end{equation}
The \SIC\ for these kets is analogous to the standard \SIC\ for qutrits
\cite{dang,stacey}.
The nine POVM elements, or probability operators, are proportional to
one-dimensional projectors,
\begin{equation}\label{SIC-1}
  \Pi_j=\frac{1}{3}\ket{v_j}\bra{v_j} \quad\text{for}\quad j=1,2,\ldots,9
\end{equation}
with
\begin{align}\label{SIC-2}
  &\mathrel{\phantom{=}}
  \column[ccccc]{\ket{v_1}&\ket{v_2}&\cdots&\ket{v_8}&\ket{v_9}}
  \nonumber\\
  &\repr\frac{1}{\sqrt{2}}\column[ccccccccc]
  { 0&-1&1&0&-\omega&1&0&-1&\omega\\
    1&0&-1&1&0&-\omega&\omega&0&-1\\
  -1&1&0&-\omega&1&0&-1&\omega&0}\,,
\end{align}
where $\omega=\mathrm{e}^{\mathrm{i}2\pi/3}$ is the basic cubit root of unity.
The sum of the elements is the projector on the triplet sector,
\begin{equation}\label{SIC-3}
  \sum_{j=1}^9\Pi_j
  =\frac{1}{4}\Bigl(3I_4+\vsigma^{(1)}\cdot\vsigma^{(2)}\Bigr)
  =I_{\mathrm{tp}}\,.
\end{equation}
After supplementing the triplet \SIC\ with the projector on the singlet,
\begin{equation}\label{SIC-4}
  \Pi_0=\frac{1}{4}\Bigl(I_4-\vsigma^{(1)}\cdot\vsigma^{(2)}\Bigr)
  =I_{\mathrm{sg}}\,,
\end{equation}
we have a proper POVM for the qubit pair.
Note that the rank of $I_{\mathrm{tp}}$ is three and that of $I_{\mathrm{sg}}$
is one.

Since
\begin{align}
  &\tr{\Pi_j\Pi_k}=\frac{1}{36}+\frac{1}{12}\delta_{jk}\,,\nonumber\\
  &\tr{\vphantom{\big|}\Pi_j(12\Pi_k-I_{\mathrm{tp}})}=\delta_{jk}
\end{align}
for ${j,k=1,2,\ldots,9}$,
we reconstruct any $\rho$ in the triplet sector from its \SIC\ probabilities
\begin{equation}\label{SICprob}
  q_j=\tr{\Pi_j\vts\rho}
\end{equation}
in accordance with
\begin{equation}
  \rho=\sum_{j=1}^9q_j(12\vts\Pi_j-I_{\mathrm{tp}})\,.
\end{equation}
In addition to being nonnegative and having unit sum, the nine probabilities
$q_j$ are subject to constraints that follow from ${\rho\geq0}$.
In particular, there is the purity constraint 
${\frac{1}{3}\leq\tr{\rho^2}\leq1}$, that is,
\begin{equation}\label{purity}
  \frac{1}{9}\leq\sum_{j=1}^9q_j^2\leq\frac{1}{6}\,.
\end{equation}
Moreover, in the present context of \Eq{genrho}, we note that these
separable rank-two triplet states make up a five-dimensional nonconvex set in
the eight-dimensional triplet sector and that restricts the permissible
probabilities stringently. 
We do not know, however, how to state these restrictions as explicit
constraints obeyed by the~$q_j$s.

\subsection{Simulated data}
For different choices of $\ket{\psi_0}$, $\ket{\psi_1}$, $p_{0}$,
and $p_{1}$, the \SIC\ was used to produce simulated measurement data;
in particular, it gave us a probability distribution from which we then
sampled.
The data are the counts $n_1,n_2,\dots,n_9$ for the nine different outcomes of 
the simulated measurement.
Two different methods were then used to find the density matrix from the data,
linear inversion and maximum likelihood.
We first used linear inversion to produce an empirical density matrix from the
data, that is, we took the relative frequencies as estimates of the
probabilities,
\begin{equation}\label{SIClin}
  \rho^{(\textsc{li})}=\sum_{j=1}^9\frac{n_j}{N}(12\vts\Pi_j-I_{\mathrm{tp}})
  \quad\text{with}\quad N=\sum_{j=1}^9n_j\,.
\end{equation}
With this in hand, we examined two methods to find the states and
probabilities.

For the first, we found the eigenvector of the empirical $3\times3$ density
matrix with the smallest eigenvalue and identified it with $\ket{\xi}$.
This ket was used to create the quadratic equation in Eq.~(\ref{quad}) and
find the states.
Once we know the states and the empirical density matrix, it is
straightforward to find the probabilities.
For the second method, we used the empirical density matrix to find the vector
$\vec{s}$ and the dyad $\vec{C}$, which then allowed us to find the vectors
$\vec{a}$, $\vec{b}$, and their probabilities.

\begin{figure}
  \centering
  \includegraphics[viewport=55 464 297 788,clip=]{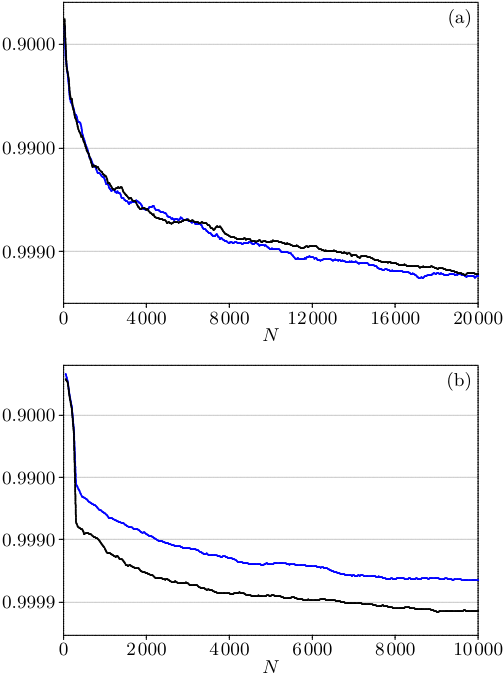}  
  \caption{\label{fig:SIC1}%
    Results from simulated data for the \SIC\ averaged over fifty runs.
    The graphs show the fidelities of the states estimated by linear inversion
    versus the number of detected pairs.
    The black curves graph
    $\magn{\braket[\big]{\smash{\psi_0^{\vphantom{(\textsc{li})}}}}
      {\smash{\psi_0^{(\textsc{li})}}}}^2$
    and the blue curves graph
    $\magn{\braket[\big]{\smash{\psi_1^{\vphantom{(\textsc{li})}}}}
      {\smash{\psi_1^{(\textsc{li})}}}}^2$.    
    Plot \textbf{(a)} is for the states
    ${\ket{\psi_0} = \ket{0}}$ and
    ${\ket{\psi_{1}} = (1/\sqrt{2})\bigl(\ket{0}+\ket{1}\bigr)}$
    and the probabilities ${p_{0}=p_{1}=1/2}$.
    In the simulation, we detect up to $20\ts000$ pairs and infer the states
    by using \Eq{quad}.
    Plot \textbf{(b)} is for the states
    ${\ket{\psi_0} = \ket{0}}$ and
    ${\ket{\psi_{1}}=\ket{1}}$ and the probabilities ${p_{0}= 0.75}$ and
    ${p_{1}= 0.25}$.
    We learn the states from \Eq{ab} for up to $10\ts000$ detected pairs and
    observe that the state with the higher probability converges faster. }
\end{figure}

In both cases, we then computed the fidelities of the states from the
simulation with the original states that were used to produce the data, and
these were plotted versus the number of pairs received.
The results for two cases are shown in Fig.~\ref{fig:SIC1}.
We used  Eq.~(\ref{quad}) for the plots in Fig.~\ref{fig:SIC1}(a), while
those in Fig.~\ref{fig:SIC1}(b) resulted from Eq.~(\ref{ab}).

The rate of convergence depends on the overlap of $\ket{\psi_0}$ and
$\ket{\psi_1}$; in particular, the larger the overlap the slower the
convergence.
In the case in which one is using the ket $\ket{\xi}$ to find the two states,
a large overlap between $\ket{\psi_0}$ and $\ket{\psi_1}$ will lead to a
$3\times3$ density matrix $\rho$ with one large and one small eigenvalue as
well as the eigenvalue~$0$ corresponding to the eigenvector $\ket{\xi}$, and
that can lead to problems.
The ket $\ket{\xi}$ satisfies ${\rho \ket{\xi} = 0}$, but the estimated density
matrix $\rho^{(\textsc{li})}$ is not in the five-dimensional space of physical
$\rho\vts$s since the relative frequencies $n_j/N$ do not obey the constraints
that restrict the probabilities~$q_j$.
Rather than the exact eigenvalue $0$ for $\ket{\xi}$, $\rho^{(\textsc{li})}$ has an
eigenvalue $\approx0$ that can be positive or negative
(while $\rho^{(\textsc{li})}$ is hermitian, nothing ensures
${\rho^{(\textsc{li})}\geq0}$). 
If, then, the overlap of $\ket{\psi_0}$ and $\ket{\psi_1}$ is very large,
$\rho$ has a small positive eigenvalue and it can be difficult to distinguish
the corresponding approximate eigenvalue of $\rho^{(\textsc{li})}$ from the
near-zero eigenvalue of $\ket{\xi}$.

For the maximum likelihood method we parameterize the density matrix
suitably. 
The two states are expressed as
${\ket{\psi_{j}} = \cos(\theta_{j})\ket{0} +
  \mathrm{e}^{\mathrm{i}\phi_{j}}\sin(\theta_{j})\ket{1}}$ for ${j=0,1}$,
so that
\begin{align}
  a_x+\mathrm{i}a_y&=\sin(2\theta_1)\,\mathrm{e}^{\mathrm{i}\phi_1}\,,
  &a_z&=\cos(2\theta_1)\,,\nonumber\\
  b_x+\mathrm{i}b_y&=\sin(2\theta_2)\,\mathrm{e}^{\mathrm{i}\phi_2}\,,
  &b_z&=\cos(2\theta_2)\,,
\end{align}
and the probabilities in terms of an angle $\alpha$,
${p_{0}=\cos(\alpha)^2}$ and ${p_{1}=\sin(\alpha)^2}$.
Then, the two-qubit density matrix of \Eq{genrho} is a function
of ${\vtheta=(\theta_{0},\phi_{0},\theta_{1},\phi_{1},\alpha)}$, the five
parameters that specify the qubit states and the probabilities, 
$\rho(\vtheta)$.
We use this density matrix to find the nine probabilities of \Eq{SICprob} as
functions of $\vtheta$, which enter the likelihood function \cite{combfac} 
\begin{equation}
  L(\vtheta)
  = \prod_{j=1}^{9} q_{j}(\vtheta)^{n_j}
\end{equation}
for the given simulated data.
For the purpose of finding the maximum of $L(\vtheta)$, we use a covariance
matrix adaptation evolution strategy (CMA-ES), adapted from
\cite{Hansen2016arXiv}, to find the minimum of
$-N^{-1}\log\bigl(L(\vtheta)\bigr)$. 
Once we have done this, we know the maximum likelihood estimates of both
states and both probabilities, which tell us $\rho^{(\textsc{ml})}$, and no
further processing is necessary.
The parameterization in terms of the five parameters ensures that every $\rho$
in the competition is in the five-dimensional set of permissible $\rho\vts$s.
Put differently, the constraints mentioned after \Eq{purity} are obeyed by
construction. 
By contrast, the unconstrained maximization of ${\log(L)=\sum_jn_j\log(q_j)}$
yields the linear inversion estimates ${q_j=n_j/N}$ with  $\rho^{(\textsc{li})}$
outside of the set of permissible $\rho\vts$s; in this sense, linear inversion
is unconstrained likelihood maximization.
Whereas $\rho^{(\textsc{ml})}$ is assuredly in the set of separable rank-two
triplet density matrices,  $\rho^{(\textsc{li})}$ is only guaranteed to be a
hermitian, unit-trace $3\times3$ matrix; it is usually of rank three and has a
roughly 50\% chance of having a negative eigenvalue.

\begin{figure}
  \centering
  \includegraphics[viewport=64 485 295 788,clip=]{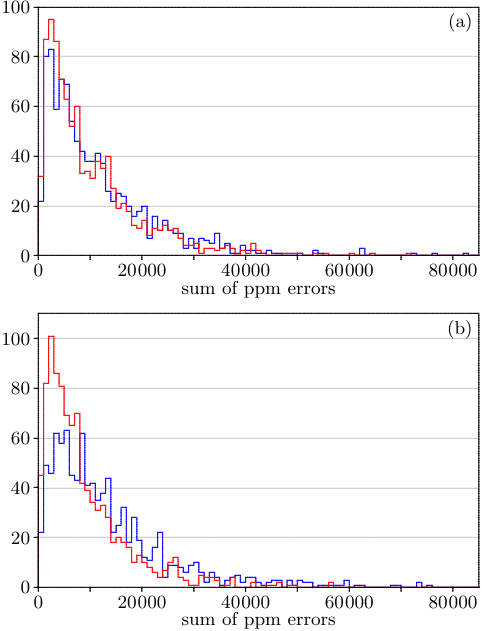}
  \caption{\label{fig:SIC2}%
    The abscissa is the sum of ppm errors for the estimated $\ket{\psi_0}$ and
    $\ket{\psi_1}$ and the histograms show the frequencies with which these
    errors occur in each bin of 1\ts000 abscissa units.
    The blue histogram is for the linear inversion method and the red histogram
    for the maximum likelihood method.
    We report results of simulated experiments 
    \textbf{(a)} for the parameters
    ${(\theta_{0},\phi_{0},\theta_{1},\phi_{1},\alpha)=%
      \left(\frac{\pi}{6},\frac{\pi}{4},\frac{2\pi}{3},\frac{\pi}{2},%
        \frac{\pi}{6}\right)}$
    and \textbf{(b)} for the parameters
    ${\left(\frac{\pi}{12},\frac{\pi}{4},\frac{5\pi}{12},\frac{\pi}{2},%
        \frac{\pi}{3}\right)}$.
    Observe that in both plots the frequency of small errors is considerably
    larger for the maximum likelihood method.}
\end{figure}

Nevertheless, linear inversion yields reasonable results as demonstrated by
the plots in Fig.~\ref{fig:SIC1}.
This is so because, although $\rho^{(\textsc{li})}$ is not in the
five-dimensional physical set, it is very close to the actual $\rho$, as
measured by a proper distance in the eight-dimensional convex
space of hermitian unit-trace $\rho\vts$s, when $N$ is sufficiently large. 

We observed that the maximum likelihood method produced higher fidelities
than the linear inversion method for a given number $N$ of data of simulated
detection events.
Here are two examples comparing the results of linear inversion and maximum
likelihood. 
For the choice ${\vtheta=
  \left(\frac{\pi}{6},\frac{\pi}{4},\frac{2\pi}{3},\frac{\pi}{2},%
    \frac{\pi}{6}\right)}$,
we simulated $750$ experiments with ${N=1\ts000}$ each.
Figure~\ref{fig:SIC2}(a) shows the sum of errors (one minus the fidelity)
for $\ket{\psi_0}$ and $\ket{\psi_1}$ expressed in parts per million.
We find that the maximum likelihood method consistently outperforms the linear
inversion method, with average errors of $10\ts546\,$ppm and $11\ts852\,$ppm,
respectively. 

The parameters
${\left(\frac{\pi}{12},\frac{\pi}{4},\frac{5\pi}{12},\frac{\pi}{2},%
    \frac{\pi}{3}\right)}$
are used in the example of Fig.~\ref{fig:SIC2}(b).
In this case, $1\ts000$ simulated experiments each with $N=1\,000$ were
conducted.
The average error for linear inversion was $15\ts760\,$ppm and for maximum
likelihood $9\ts283\,$ppm.  

In conclusion, the maximum likelihood method is more computationally
demanding than the linear inversion method, due to the maximization, but
produces better results.
It also directly finds the two unknown states because, once the optimal
parameters are known, then so are the states, whereas the linear inversion
first yields a density matrix, and then that has to be diagonalized to find
the states.
The simulations of Figs.~\ref{fig:SIC1} and \ref{fig:SIC2} confirm that these
two point estimators are consistent, that is, ${\rho^{(\textsc{li})}\to\rho}$
and ${\rho^{(\textsc{ml})}\to\rho}$ as ${N\to\infty}$;
a quantitative statement about this is provided in
Sec.~\ref{sec:tetra-B} in the context of the tetrahedron POVM,
see Eqs.~(\ref{plaus6}) and (\ref{plaus7}),
with additional details in the Appendix.

\begin{figure}
  \centering
  \includegraphics[viewport=65 753 293 789,clip=]{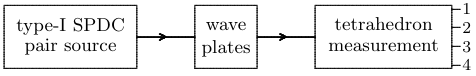}
  \caption{\label{fig:tetra}%
    A pump laser (not indicated), pulsed or continuous, illuminates a
    crystal for type-I spontaneous parametric down conversion (SPDC),
    which acts as a
    source of pairs of photons that propagate in the same direction and have
    the same polarization---both vertically polarized, say.
    A set of wave plates is used to change the polarization from vertical to
    any other kind.
    We switch at random between the two settings of the wave plates that
    correspond to the polarizations specified by the Bloch vectors $\vec{a}$
    and $\vec{b}$ with the respective probabilities $p_0$ and $p_1$ for the
    next pair. 
    The photons are detected by a tetrahedron measurement
    \cite{REK:04,ETGN:05,LSLLK:06a,LSLLK:06b}, the \SIC\ for single qubits,
    where we either register both photons in one exit port (four cases) or get
    a coincidence between two different exit ports (six cases).}
\end{figure}

\section{Tetrahedron POVM}\label{sec:tetra}
\subsection{The measurement}\label{sec:tetra-A}
A tomography experiment that could be performed with existing technology is
sketched in Fig.~\ref{fig:tetra}.
The elements of the single-qubit tetrahedron POVM are
\begin{equation}
  \Pi_j^{(\textsf{4})}
  =\frac{1}{4}(I_2+\vec{t}_j\cdot\vsigma)\quad\text{with}\quad j=1,2,3,4\,;
\end{equation}
the four tetrahedron vectors have the properties
\begin{equation}
  \vec{t}_j\cdot\vec{t}_k=\frac{4}{3}\delta_{jk}-\frac{1}{3}\,,\quad
  \sum_{k=1}^4\vec{t}_k=0\,,\quad
  \sum_{k=1}^4\vec{t}_k\vec{t}_k=\frac{4}{3}\vec{1}\,,
\end{equation}
where $\vec{1}$ is the unit dyad.
Accordingly, the four POVM elements for detecting both photons at the
same exit are
\begin{align}\label{T-1}
  \Pi_{k}^{(\text{s})}
  &=\Pi_k^{(\textsf{4})}\otimes\Pi_k^{(\textsf{4})}
    \nonumber\\
  &=\frac{1}{16}\Bigl(I_4+\vec{t}_k\cdot(\vsigma^{(1)}+\vsigma^{(2)})
    +\vsigma^{(1)}\cdot\vec{t}_k\vec{t}_k\cdot\vsigma^{(2)}\Bigr)
\end{align}
for $k=1,2,3,4$
and the six elements for the coincidence counts are
\begin{align}\label{T-2}
  \Pi_{jk}^{(\text{c})}
  &=\Pi_j^{(\textsf{4})}\otimes\Pi_k^{(\textsf{4})}
    +\Pi_k^{(\textsf{4})}\otimes\Pi_j^{(\textsf{4})}\nonumber\\
   &=\frac{1}{16}\Bigl(2I_4+(\vec{t}_j+\vec{t}_k)
         \cdot(\vsigma^{(1)}+\vsigma^{(2)})
    \nonumber\\
  &\phantom{=\frac{1}{16}\Bigl(}
    +\vsigma^{(1)}\cdot(\vec{t}_j\vec{t}_k
    +\vec{t}_k\vec{t}_j)\cdot\vsigma^{(2)}\Bigr) 
\end{align}
for ${1\leq j<k\leq4}$.
Clearly, the \SIC\ of Eqs.~(\ref{SIC-1})--(\ref{SIC-4}) is not of the kind
realized by the elements in Eqs.~(\ref{T-1}) and (\ref{T-2}) or
any other product POVM. 
In particular, we have nonzero probabilities for qubit pairs in the singlet
state,
\begin{equation}
  \tr{\Pi_{k}^{(\text{s})}I_{\mathrm{sg}}}=0\,,\qquad
  \tr{\Pi_{jk}^{(\text{c})}I_{\mathrm{sg}}}=\frac{1}{6}\,.
\end{equation}

The ten-outcome measurement with the POVM elements in Eqs.~(\ref{T-1})
and (\ref{T-2}) is tomographically complete in the space of density operators
that are convex sums of the singlet state and any mixed state in the triplet
sector, that is,
\begin{equation}
  \rho=\frac{1}{4}\Bigl(I_4
        +\vec{s}\cdot\bigl(\vsigma^{(1)}+\vsigma^{(2)}\bigr)
        +\vsigma^{(1)}\cdot\vec{C}\cdot\vsigma^{(2)}\Bigr)
\end{equation}
with ${C_{jk}=C_{kj}}$.
The singlet in \Eq{SIC-4} is of this kind and so is, of course,  the state
emitted by the source with $\vec{s}$ and $\vec{C}$ of the particular forms
in \Eq{gensC}.
We recall the single-qubit identity
\begin{equation}
  \qquad\tr{\Pi_j^{(\textsf{4})}R_k^{(\textsf{4})}}=\delta_{jk}
  \quad\text{with}\quad
    R_k^{(\textsf{4})}
    =\frac{1}{2}(I_2+3\vec{t}_k\cdot\vsigma)\qquad
\end{equation}
and recognize that the reconstruction operators $R_{k}^{(\text{s})}$ and
$R_{jk}^{(\text{c})}$ that are defined by
\begin{equation}
  \begin{array}[b]{@{}rcl@{\enskip}rcl@{}}
    \tr{\Pi_{k}^{(\text{s})}R_{k'}^{(\text{s})}}
    &=&\delta_{k\vts k'}\,,
    &
    \tr{\Pi_{k}^{(\text{s})}R_{j'k'}^{(\text{c})}}&=&0\,,\\[2ex]
    \tr{\Pi_{jk}^{(\text{c})}R_{k'}^{(\text{s})}}
    &=&0\,,
    &
      \tr{\Pi_{jk}^{(\text{c})}R_{j'k'}^{(\text{c})}}
    &=&\delta_{j\vts j'}\,\delta_{k\vts k'}
  \end{array}
\end{equation}
are
\begin{align}
  R_{k}^{(\text{s})}
  &=R_k^{(\textsf{4})}\otimes R_k^{(\textsf{4})}\nonumber\\
  &=\frac{1}{4}\Bigl(I_4
    +3\vec{t}_k\cdot\bigl(\vsigma^{(1)}+\vsigma^{(2)}\bigr)
    +9\ts\vsigma^{(1)}\cdot\vec{t}_k\vec{t}_k\cdot\vsigma^{(2)}\Bigr)\,,
    \nonumber\\
  R_{jk}^{(\text{c})}
  &=\frac{1}{2}R_j^{(\textsf{4})}\otimes R_k^{(\textsf{4})}
    +\frac{1}{2}R_k^{(\textsf{4})}\otimes R_j^{(\textsf{4})}
    \nonumber\\
  &=\frac{1}{8}\Bigl(2I_4+3(\vec{t}_j+\vec{t}_k)
    \cdot\bigl(\vsigma^{(1)}+\vsigma^{(2)}\bigr)\nonumber\\
  &\phantom{=\frac{1}{8}\Bigl(}
  +9\ts\vsigma^{(1)}\cdot(\vec{t}_j\vec{t}_k+\vec{t}_k\vec{t}_j)
    \cdot\vsigma^{(2)}\Bigr)\,.  
\end{align}
Accordingly,
\begin{equation}\label{reconstruct}
  \rho=\sum_{k=1}^4q_k^{(\text{s})}R_{k}^{(\text{s})}
       +\sum_{j<k}q_{jk}^{(\text{c})}R_{jk}^{(\text{c})}
\end{equation}
reconstructs $\rho$ from the probabilities
\begin{equation}\label{tetraprob}
  q_k^{(\text{s})}=\tr{\Pi_{k}^{(\text{s})}\rho}\,,\quad
  q_{jk}^{(\text{c})}=\tr{\Pi_{jk}^{(\text{c})}\rho}\,,
\end{equation}
which means
\begin{align}
  \vec{s}&=3\sum_{k=1}^4q_k^{(\text{s})}\vec{t}_k
           +\frac{3}{2}\sum_{j<k}q_{jk}^{(\text{c})}(\vec{t}_j+\vec{t}_k)\,,
           \nonumber\\
  \vec{C}&=9\sum_{k=1}^4q_k^{(\text{s})}\vec{t}_k\vec{t}_k
              +\frac{9}{2}\sum_{j<k}q_{jk}^{(\text{c})}
              (\vec{t}_j\vec{t}_k+\vec{t}_k\vec{t}_j)
\end{align}
for the Bloch vector $\vec{s}$ and the dyad $\vec{C}$.
For the $\vec{s}$ and $\vec{C}$ in \Eq{gensC}, the sum rules
\begin{equation}\label{sumrules}
  \sum_{k=1}^4 q_k^{(\text{s})}=\frac{1}{3}\,,\quad
  \sum_{j=1}^3\sum_{k=j+1}^4q_{jk}^{(\text{c})}=\frac{2}{3}
\end{equation}
apply.

The data are the counts $n_k^{(\text{s})}$ and $n_{jk}^{(\text{c})}$ of the
ten different outcomes with the total count of detected pairs
\begin{equation}
  N=\sum_{k=1}^4n_k^{(\text{s})}+\sum_{j<k}n_{jk}^{(\text{c})}\,.
\end{equation}
The law of large numbers states that the
relative frequencies ${n_k^{(\text{s})}\big/N}$ and
${n_{jk}^{(\text{c})}\big/N}$ approximate the respective
probabilities when ${N\gg1}$ and, therefore, we get an estimate for $\rho$ by
replacing the probabilities in \Eq{reconstruct} by the relative frequencies
\begin{equation}
  \rho\cong\rho^{(\textsc{li})}
  =\frac{1}{N}\sum_{k=1}^4n_k^{(\text{s})}R_{k}^{(\text{s})}
       +\frac{1}{N}\sum_{j<k}n_{jk}^{(\text{c})}R_{jk}^{(\text{c})}\,,
\end{equation}
the analog of \Eq{SIClin}.
As discussed above, this linear inversion is problematic because the relative
frequencies do not obey the constraints that apply to the probabilities, such
as the sum rules in \Eq{sumrules}.
For instance, while ${\tr{I_{\text{sg}}\rho}=0}$ for the actual
$\rho$, we have
\begin{equation}
  \tr{I_{\text{sg}}\rho^{(\textsc{li})}}
  =-\frac{2}{N}\sum_{k=1}^4n_k^{(\text{s})}
    +\frac{1}{N}\sum_{j<k}n_{jk}^{(\text{c})}\,,
\end{equation}
which is almost always nonzero and negative half the time.
While we can improve matters a bit by removing the singlet component from
$\rho^{(\textsc{li})}$,  
the resulting mixed triplet state is almost always \emph{not} a rank-two
separable state and can have a negative eigenvalue in the triplet sector.

\subsection{Plausible states}\label{sec:tetra-B}
Rather than merely finding the point estimators of the linear inversion method,
the maximum likelihood method, or yet other methods, we identify the plausible 
region in the parameter space \cite[Sec.\ 4.5.2]{Evans},
the set of all separable rank-two triplet
states that are supported by the data.
Every $\rho$ in the plausible region is an acceptable point estimator;
additional criteria, beyond what the data tell us, would be needed for
selecting a particular one. 
While the linear inversion estimator $\rho^{(\textsc{li})}$ is usually
improper, the maximum likelihood estimator $\rho^{(\textsc{ml})}$ is always
plausible, because the plausible region is one of the optimal
error regions, which happen to be regions around the maximum likelihood
estimator \cite{SNSLE:13}.

If we denote the prior probability element of the vicinity of $\rho(\vtheta)$
by $(\mathrm{d}\vtheta)$, the posterior probability element is
\begin{equation}\label{plaus1}
  (\mathrm{d}\vtheta)_{\text{post}}=\frac{(\mathrm{d}\vtheta)\,L(\vtheta)}
  {\displaystyle\int(\mathrm{d}\vtheta')\,L(\vtheta')}\,.
\end{equation}
The data give evidence in favor of $\rho(\vtheta)$ if
${(\mathrm{d}\vtheta)_{\text{post}}>(\mathrm{d}\vtheta)}$ and evidence against
$\rho(\vtheta)$ if ${(\mathrm{d}\vtheta)_{\text{post}}<(\mathrm{d}\vtheta)}$;
the data are neutral when
${(\mathrm{d}\vtheta)_{\text{post}}=(\mathrm{d}\vtheta)}$. 
The plausible region is composed of all $\rho(\vtheta)$s with evidence in their
favor, that is,
\begin{equation}
  \parbox[b]{16.5em}{
  the density matrix $\rho(\vtheta)$ is plausible if
  $$L(\vtheta)>\int(\mathrm{d}\vtheta')\,L(\vtheta')$$
  and only then.}
\end{equation}
This is an application of the principle of evidence \cite[Sec.~4.2]{Evans};
another application to quantum data is reported in \cite{GLEE:19}.

For the given data, we find the maximum likelihood estimator
${\rho^{(\textsc{ml})}=\rho\bigl(\vtheta^{(\textsc{ml})}\bigr)}$
and then the number
\begin{equation}\label{plaus3}
  \lambda_{\text{pl}}=\frac{1}{L\bigl(\vtheta^{(\textsc{ml})}\bigr)}
  \int(\mathrm{d}\vtheta)\,L(\vtheta)<1\,,
\end{equation}
so that $\rho(\vtheta)$ is plausible if
\begin{equation}
  \lambda(\vtheta)
  =\frac{L(\vtheta)}{L\bigl(\vtheta^{(\textsc{ml})}\bigr)}>\lambda_{\text{pl}}\,.
\end{equation}
The data give strong evidence 
if the prior content (``size'') of the plausible
region is small and its posterior content (``credibility'') is large;
these are
\begin{align}\label{plaus5}
  \text{size:}\quad
  &s^{\,}_{\text{pl}}=\int(\mathrm{d}\vtheta)\,
    \Chi\Bigl(\lambda(\vtheta)>\lambda_{\text{pl}}\Bigr)\,,
  \nonumber\\
  \text{credibility:}\quad
  &c^{\,}_{\text{pl}}=\int(\mathrm{d}\vtheta)_{\text{post}}\,
    \Chi\Bigl(\lambda(\vtheta)>\lambda_{\text{pl}}\Bigr)\,,
\end{align}
where $\Chi(\text{A})=1$ if the statement $\text{A}$ is true and
$\Chi(\text{A})=0$ if it is false.

When there are many data so that the total count $N$ of detection events is
large and the law of large numbers (central limit theorem) is applicable,
the $N$ dependence of $\lambda_{\text{pl}}$, $s^{\,}_{\text{pl}}$,  and
$c^{\,}_{\text{pl}}$ is
given by \cite[Sec.\ 7.4]{LQSE}
\begin{equation}\label{plaus6}
  \lambda_{\text{pl}}  \propto N^{-5/2}
\end{equation}
with the proportionality factor depending on the relative frequencies and
\begin{align}\label{plaus7} 
  s_{\text{pl}}^{\,}
  &\cong\lambda_{\text{pl}}
  \frac{\log(1/\lambda_{\text{pl}})^{5/2}}
    {\rule{0pt}{10pt}(5/2)!}
    \propto N^{-5/2}\log(N)^{5/2} \,,\nonumber\\
  1-c_{\text{pl}}^{\,}
  &\cong\int_{0}^{\lambda_{\text{pl}}}\mathrm{d}\lambda\,
    \frac{\log(1/\lambda)^{3/2}}{(3/2)!}
    \nonumber\\
  &=s_{\text{pl}}^{\,}\biggl(\frac{5}{2}\log(1/\lambda_{\text{pl}})^{-1}
    +\frac{15}{4}\log(1/\lambda_{\text{pl}})^{-2}\biggr)
    \nonumber\\
  &\mathrel{\phantom{=}}{}
    +\mathrm{erfc}\Bigl(\log(1/\lambda_{\text{pl}})^{1/2}\Bigr)
  \nonumber\\
   &\cong\lambda_{\text{pl}}
  \frac{\log(1/\lambda_{\text{pl}})^{3/2}}
    {\rule{0pt}{10pt}(3/2)!}
    \propto N^{-5/2}\log(N)^{3/2} \,.
\end{align}
Accordingly, the size of the plausible region shrinks with growing $N$ and so
does the gap between its credibility and unity.
When there are many data, the plausible region is very small and has very
large credibility---very small prior and very large posterior probability.

The five-dimensional integrals in Eqs.~(\ref{plaus1})--(\ref{plaus5}) are
evaluated with Monte Carlo methods as follows \cite[Sec.\ 8.2]{LQSE}.
We draw a large sample of $\vtheta$s from the prior distribution and store the
corresponding $\lambda(\vtheta)$ values:
$\lambda^{(1)},\lambda^{(2)},\dots,\lambda^{(m)},\dots,\lambda^{(M)}$.
Then,
\begin{align}
  \lambda_{\text{pl}}&\cong\frac{1}{M}\sum_{m=1}^M\lambda^{(m)}\,,\nonumber\\
  s^{\,}_{\text{pl}}&\cong\frac{1}{M}\sum_{m=1}^M
                 \Chi\Bigl(\lambda^{(m)}>\lambda_{\text{pl}}\Bigr)\,,\nonumber\\
  c^{\,}_{\text{pl}}&\cong\frac{1}{M\lambda_{\text{pl}}}\sum_{m=1}^M\lambda^{(m)}
                 \Chi\Bigl(\lambda^{(m)}>\lambda_{\text{pl}}\Bigr)
\end{align}
are approximate values for $\lambda_{\text{pl}}$, $s^{\,}_{\text{pl}}$,  and
$c^{\,}_{\text{pl}}$ with a sampling error ${\propto 1/\sqrt{M}}$. 
As it is usually CPU-cheap to draw from the prior distribution, we can easily
have samples that are so large that the sampling error is of no concern.

\begin{figure}
  \centering
  \includegraphics[viewport=56 650 296 790,clip=]{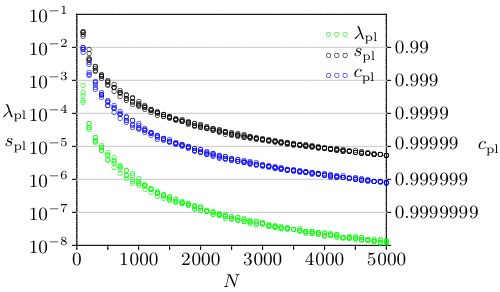}
  \caption{\label{fig:TetraSim-1}%
   The values of $\lambda_{\text{pl}}$, $s^{\,}_{\text{pl}}$,  and
   $c^{\,}_{\text{pl}}$ as a function of $N$ for five simulated experiments
   with up to $5\vts000$ detected pairs each.
   We observe that there is little variation between the experiments and that a
   few thousand detected pairs is enough to ensure a small size and large
   credibility of the plausible region.}
\end{figure}

\subsection{Simulated data}\label{sec:tetra-C}
We simulated five runs of the experiment of Fig.~\ref{fig:tetra} for
the Bloch vectors specified by
\begin{align}
  \column{\vec{a}\cdot\vec{t}_1\\
          \vec{a}\cdot\vec{t}_2\\
          \vec{a}\cdot\vec{t}_3\\
          \vec{a}\cdot\vec{t}_4}
       & =\frac{1}{\sqrt{27}}
        \column[r]{-3\\-1\\-1\\5}\,,\nonumber\\
  \column{\vec{b}\cdot\vec{t}_1\\
          \vec{b}\cdot\vec{t}_2\\
          \vec{b}\cdot\vec{t}_3\\
          \vec{b}\cdot\vec{t}_4}
        &=\frac{1}{\sqrt{105}}
        \column[r]{-3\\9\\1\\-7}
\end{align}
and the probabilities ${p_0=0.37}$, ${p_1=0.63}$.
Figure \ref{fig:TetraSim-1} shows $\lambda_{\text{pl}}$, $s^{\,}_{\text{pl}}$,  and
$c^{\,}_{\text{pl}}$ for ${N=100}$, $200$, $300$, \dots, $5\vts000$ for the
five runs.
There is very little variation between the runs.
The plausible region has a credibility that exceeds $0.999\ts9$
for ${N=1\ts000}$, $0.999\vts99$ for ${N=2\ts000}$, and $0.999\ts999$ for
${N=5\ts000}$, while the size is less than $10^{-3}$, $10^{-4}$, and
$10^{-5}$, respectively.
It follows that a few thousand detected pairs are quite enough
to acquire very strong evidence in favor of a tiny subset of the separable
rank-two triplet states and against all others.
We verified that the $\rho(\vtheta)$ used for the simulation
is in the plausible region for ${N=100}$, $200$, \dots, $5\ts000$, as
it is expected to be:
For ${N=100}$ already, the credibility is $0.99$ so that the
true state has an outside probability of only 1\%.
In passing, we note that the Monte Carlo integration requires ridiculously
large samples for ${N>5\ts000}$ because the peak of the likelihood function
around the maximum likelihood estimator is then extremely narrow.

We used a sample with ${M=2\times10^9}$ entries for the Monte Carlo
integration. 
It was drawn from the prior probability distribution that has the Bloch
vectors $\vec{a}$ and $\vec{b}$ independently uniformly (isotropically)
distributed on the unit sphere and the probability parameter $\alpha$
uniformly distributed between $0$ and $\pi/2$.
This prior correctly reflects our complete ignorance about $\vec{a}$,
$\vec{b}$, $p_0$, and $p_1$ before we take data, whereby we recall Wootters's
insight \cite{W:81} that the Jeffreys prior \cite{J:46} is most natural for
the probabilities.

\begin{figure}
  \centering
\includegraphics[viewport=72 520 278 790,clip]{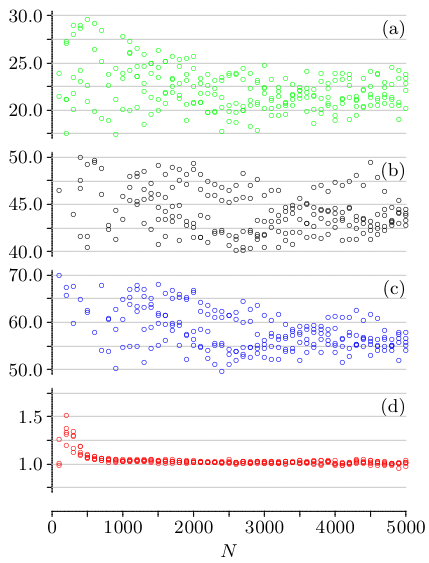}  
  \caption{\label{fig:TetraSim-2}%
    An illustration of the large-$N$ approximations
    in Eqs.\ (\ref{plaus6}) and (\ref{plaus7}); see text.}
\end{figure}

Let us exploit the simulated data for a check of the large-$N$ approximations
in Eqs.\ (\ref{plaus6}) and (\ref{plaus7}).
Figure \ref{fig:TetraSim-2} summarizes the data of the five simulated
experiments. 
Plot (a) shows the values of $N^{5/2}\lambda_{\text{pl}}\,$;
plot (b) shows those of $N^{5/2}\log(N)^{-5/2}s_{\text{pl}}^{\,}\,$;
plot (c) those of $N^{5/2}\log(N)^{-3/2}\bigl(1-c_{\text{pl}}^{\,}\bigr)\,$;
and plot (d) those of
$\bigl(1-c_{\text{pl}}^{\,}\bigr)\Big/\Bigl(s_{\text{pl}}^{\,}\bigr(\cdots\bigr)
+\mathrm{erfc}(\cdot)\Bigr)$
with the approximation in \Eq{plaus7}.
It appears that a few thousand detected pairs are sufficiently many to reach
the asymptotic regime, where the ratio (d) is unity,
and plots (a)--(c) show fluctuations around an
$N$-independent value.
We observe that the ratio in (d) is least
sensitive to the fluctuations in the relative frequencies.

\subsection{Imperfections in real-life experiments}\label{sec:tetra-D}
The above discussion of the tetrahedron POVM assumes an ideal realization,
whereas an actual experiment will have imperfections.
In particular, there are deviations from the perfect tetrahedron geometry,
the detectors at the four outputs have nonideal detection
efficiencies, and the optical elements between the source and the detectors
will absorb or deflect a small fraction of the photons.
As a consequence, there will be cases where only one of the two photons is
detected or both escape detection.
The ten POVM elements of Eqs.\ (\ref{T-1}) and (\ref{T-2}) are then
modified accordingly and supplemented by five additional elements that
account for the single-photon detection events and the null event.
It is well-known how to do all that (see, for example, \cite{GLEE:19,SSNE:19})
and we shall address these matters in due course, namely when experimental
data will be available for evaluation.

\section{Conclusions}
To conclude, we considered the following problem.  One receives a sequence of
qubits, where each qubit is in one of two unknown states, and the goal is to
determine the states.
Solving this requires some extra information.
We have shown that providing
additional quantum information, in the form of an additional copy of the
state, so that one is receiving a sequence of pairs of states instead of
individual states, allows one to determine the unknown quantum states as well
as their probabilities of occurrence.

Proper quantum state tomography provides the data from which one can learn the
unknown states and probabilities.
We have analyzed the \SIC\ and showed that the linear inversion method and the
maximum likelihood method can be used, the former with caution.
We have also proposed a tomography experiment that can be realized with
available technology and have demonstrated that a few thousand detected pairs
are enough to locate the states and probabilities within a very small region
with very high probability.

\acknowledgments 
One of us (MH) would like to thank Sarah Croke for a useful conversation.
This research was supported by NSF grant FET-2106447.
This project is supported by the National Research Foundation, Singapore
through the National Quantum Office, hosted in A*STAR, under its Centre for
Quantum Technologies Funding Initiative (S24Q2d0009).

\appendix
\section{\SIC\ and linear inversion}\label{appA}
When the qubit pairs are measured by the \SIC\ of
Eqs.~(\ref{SIC-1})--(\ref{SIC-4}), there are no counts associated with $\Pi_0$
so that
\begin{equation}\label{likelihood}
  L(\lst{n}|\lst{q})
  = \frac{N!}{\prod_{j=1}^{9} n_{j}!} \prod_{k=1}^{9} q_{k}^{n_{k}} .
\end{equation}
is the likelihood of the data ${\lst{n}=(n_1,n_2,\dots,n_9)}$, the list of
detection event counts, given the list of detection probabilities
${\lst{q}=(q_1,q_2,\dots,q_9)}$; the combinatorial factor is that for a
pre-chosen number $N$ of detected pairs.
The expected value of a function of the data,
\begin{equation}
  \expect[\big]{f(\lst{n})}=\sum_{\lst{n}}f(\lst{n})L(\lst{n}|\lst{q})\,,
\end{equation}
considers all thinkable data.
The generating function
\begin{equation}
  g(\lst{x})
  =\expect{\prod_{j=1}^9x_j^{n_j}}={\left(\sum_{j=1}^9x_jq_j\right)}^{\!N}
\end{equation}
provides the expected values of powers of the counts $n_j$,
in particular,
\begin{align}
  \expect{n_k}
  &=x_k\frac{\partial}{\partial x_k}g(\lst{x})\biggr|_{\text{all\ }x_j=1}
    =Nq_k\,,\nonumber\\
  \expect{n_jn_k}
  &=N(N-1)q_jq_k+\delta_{jk}Nq_k\,.
\end{align}
It follows that the linear inversion estimator in \Eq{SIClin} is unbiased,
${\expect{\rho^{(\textsc{li})}}=\rho}$.
The Hilbert--Schmidt norm of the error
${\Delta\rho=\rho^{(\textsc{li})}-\rho}$ is given by
\begin{equation}
  \bigl\|\Delta\rho\ts\bigr\|^2
  =\tr{\vphantom{\Big|}(\Delta\rho)^2}
  =12\sum_{j=1}^9\biggl(\frac{n_j}{N}-q_j\biggr)^2\,;
\end{equation}
as a consequence of the purity constraint in \Eq{purity}, the expected value
\begin{equation}\label{error}
  \expect{\bigl\|\Delta\rho\ts\bigr\|^2}
  =\frac{12}{N}{\left(1-\sum_{j=1}^9q_j^2\right)}
\end{equation}
is bounded by
${10/N\leq\expect{\bigl\|\Delta\rho\ts\bigr\|^2}\leq32/(3N)}$.

We want to use the empirical density matrix $\rho^{(\textsc{li})}$ to find the
triplet ket $\ket{\xi}$ that is orthogonal 
to $\ket{\psi_0}\otimes\ket{\psi_0}$ and $\ket{\psi_1}\otimes\ket{\psi_1}$.
We recall that
\begin{equation}
  \rho^{(\textsc{li})}=\sum_{j=0}^2r_{j}^{\,}\,\ket{u_j}\bra{u_j}
  \quad\text{with}\quad
  r^{\,}_0\leq r^{\,}_1\leq r^{\,}_2
\end{equation}
is hermitian but it has rank three almost always and the eigenvalue $r^{\,}_0$
can be negative in this spectral decomposition.
We have that ${\rho\ket{\xi}=0}$ and want to find a lower bound on the overlap
between $\ket{\xi}$ and $\ket{u_0}$ under the assumption that
${\|\Delta\rho\|=\epsilon\ll1}$, which we can ensure by choosing $N$ large
enough. 
The eigenvalues of $\rho^{(\textsc{li})}$ are then $\epsilon$-close to those
of $\rho$,
\begin{align}
  r_0&\cong0\,,\nonumber\\
  \begin{array}{l}
    r^{\,}_1 \\ r^{\,}_2
  \end{array}\biggr\}
  &\cong\frac{1}{2}
  \mp\frac{1}{2}\sqrt{(p_0-p_1)^2+4p_0p_1\magn{\braket{\psi_0}{\psi_1}}^4}\,.
\end{align}
We take for granted that $\epsilon$ is small enough that we can distinguish
between ${r_0\cong0}$ and ${r_1>0}$. 

Now,
\begin{equation}
  \magn{\bra{\xi}\Delta\rho\ket{\xi}}=
  \magn[\Bigg]{\sum_{j=0}^2r^{\,}_j\magn{\braket{u_j}{\xi}}^2}\leq\epsilon\,,
\end{equation}
where
\begin{equation}
  \magn{\braket{u_0}{\xi}}^2=1-\sum_{j=1}^2\magn{\braket{u_j}{\xi}}^2\,.
\end{equation}
Since
\begin{equation}
  \sum_{j=1}^2\magn{\braket{u_j}{\xi}}^2
  \leq\frac{1}{r^{\,}_1}\sum_{j=1}^2r^{\,}_j\magn{\braket{u_j}{\xi}}^2
  \leq\frac{\epsilon}{r^{\,}_1}
\end{equation}
if ${r^{\,}_0\geq0}$, it follows that
\begin{equation}
  \magn{\braket{u_0}{\xi}}^2\geq1-\frac{\epsilon}{r^{\,}_1}
\end{equation}
in this case.
If ${r^{\,}_0<0}$, then
\begin{align}
  \epsilon
  & \geq  -\magn{r^{\,}_{0}}\,\magn{\braket{u_0}{\xi}}^2
           +\sum_{j=1}^2r^{\,}_j\magn{\braket{u_j}{\xi}}^2 \nonumber \\
  & \geq   -\magn{r^{\,}_{0}}\,\magn{\braket{u_0}{\xi}}^2
    + r^{\,}_1 \Bigl(1- \magn{\braket{u_0}{\xi}}^2\Bigr)
\end{align}
so
\begin{equation}
  \magn{\braket{u_0}{\xi}}^2
  \geq\frac{r^{\,}_1-\epsilon}{r^{\,}_1+\magn{r^{\,}_{0}}}\,.
\end{equation}
Note that
${\bra{u_0}\Delta\rho\ket{u_0}=r^{\,}_0-\bra{u_0}\rho\ket{u_0}}$
so that $\epsilon \geq \magn{r^{\,}_{0}} + \bra{u_{0}}\rho\ket{u_{0}}
  \geq\magn{r^{\,}_0}$ in this case.
Therefore,
\begin{equation}
  \magn{\braket{u_0}{\xi}}^2
  \geq 1-\frac{\magn{r^{\,}_{0}}+\epsilon}{r^{\,}_1+\magn{r^{\,}_{0}}}
  \geq 1-\frac{2\epsilon}{r^{\,}_1+\magn{r^{\,}_{0}}} \,,
\end{equation}
and the second term is of order $\epsilon$. 

What we see from this is that if $r^{\,}_{1}$ is small---recall that
$\magn[]{r^{\,}_0}$ is assumed to be substantially smaller---
then a very small value of $\epsilon$ will be required for
$\magn{\braket{u_0}{\xi}}$ to be close to one.
We can identify $\epsilon$ with the square root of the expected value in
\Eq{error}, so this gives us an estimate for how many qubit pairs we will
have to detect to obtain a good estimate of $\ket{\xi}$.

\section{The plausible region when ${N\gg1}$}\label{appB}
The consideration of all thinkable data in
Eqs.~(\ref{likelihood})--(\ref{error}) is appropriate when planning the
experiment.
Once the experiment has been performed, we draw inference form the actual
data in conjunction with what we knew before we had the data.
This prior knowledge is reflected in the prior probability element
\begin{equation}
  (\mathrm{d}\vtheta)=\mathrm{d}\theta_0\,\mathrm{d}\phi_0\,
  \mathrm{d}\theta_1\,\mathrm{d}\phi_1\,\mathrm{d}\alpha\;w(\vtheta)\,,
\end{equation}
where $w(\vtheta)$ is the prior probability density in this parameterization.
For the list $\lst{q}$ of the ten probabilities 
of the tetrahedron POVM in \Eq{tetraprob} and the corresponding list $\lst{n}$
of counts, the fractional likelihood is
\begin{align}
  \frac{L\bigl(\lst{n}|\lst{q}(\vtheta)\bigr)}
  {L\Bigl(\lst{n}|\lst{q}\bigl(\vtheta^{(\textsc{ml})}\bigr)\Bigr)}
  &=\prod_j
  \Bigl(q_j(\vtheta)\Big/q_j\bigl(\vtheta^{(\textsc{ml})}\bigr)\Bigr)^{n_j}
  \nonumber\\
  &\cong\exp{\left(-\frac{N}{2}\sum_{j,k=1}^5
    \varepsilon_j Q_{jk}\bigl(\vtheta^{(\textsc{ml})}\bigr)\varepsilon_k\right)}\,,
\end{align}
where ${\vtheta-\vtheta^{(\textsc{ml})}=(\varepsilon_1,\dots,\varepsilon_5)}$
and $Q_{jk}\bigl(\vtheta^{(\textsc{ml})}\bigr)$
is a matrix element of a symmetric positive $5\times5$ matrix;
the approximation is valid when $N$ is so large that we can invoke the
central limit theorem.
Under these circumstances, only the vicinity of $\vtheta^{(\textsc{ml})}$
contributes substantially to the integrals in Eqs.~(\ref{plaus3}) and
(\ref{plaus5}) and, correspondingly, the prior probability element can be
replaced by
\begin{equation}
  (\mathrm{d}\vtheta)\cong
  \mathrm{d}\varepsilon_1\,\cdots\mathrm{d}\varepsilon_5
    \;w\bigl(\vtheta^{(\textsc{ml})}\bigr)\,.
\end{equation}
The large-$N$ approximations in Eqs.~(\ref{plaus6}) and
(\ref{plaus7}) follow \cite{LQSE}.

\newcommand*{\arX}[2][]{\ and e-print arXiv:\linebreak[0]#1\linebreak[0]#2}



\begin{thebibliography}{99}
\bibitem{anon}
  We would like to thank an anonymous referee for pointing this out.
\bibitem{combfac}
  We ignore the combinatorial factors that have no bearing on what follows.
\end{thebibliography}
\end{document}